# Quantum steering of electron wave function in an InAs Y-branch switch

G. M. Jones,[1] M.J. Yang,[2] Y. B. Lyanda-Geller[3] and C.H. Yang[1]

[1]Department of Electrical and Computer Engineering, University of Maryland, College Park, MD 20742

[2]Naval Research Laboratory, Washington DC 20375
[3]Department of Physics, Purdue University, West Lafayette IN 47907

*Abstract*

We report experiments on gated Y-branch switches made from InAs ballistic electron wave guides. We demonstrate that gating modifies the electron wave functions as well as their interference pattern, causing the anti-correlated, oscillatory transconductances. Such previously unexpected phenomenon provides evidence of steering the electron wave function in a multi-channel transistor structure.

Quantum effects in nanostructures provide insights into fundamental issues that cannot be addressed in atomic physics settings and offer perspectives for future applications in computing. In the regime where quantum effects dominate, electron transport exhibits fundamentally new properties. For example, when the device size becomes less than the elastic mean free path, electrons can traverse through the conductor ballistically, leading to conductance quantization. In addition, phase coherent transport plays an important role in nanometer-scale devices. Among devices exploiting these quantum effects, the Y-channel transistor is attractive on its own right. The original proposal of the Y-channel transistor, or Y branch switch (YBS)[1] came from an electron wave analogy to the fiber optic coupler. The semiconductor version of YBS has a narrow electron waveguide patterned into a "Y" configuration with one source and two drain terminals. A lateral electric field perpendicular to the direction of electric current in the source waveguide steers the injected electron wave into either of the two outputs. YBS offers several advantages as a fast switch. Based on electron wave steering, the ultimate size of the switch can be downscaled to nanometers, leading to the switching speed of devices in the THz range. Most interestingly, switching can be accomplished by a voltage of the order of $\hbar/(e\tau_t)$, where $\tau_t$ is the transit time of electrons. Then, the switching voltage for a YBS can become smaller than the thermal voltage, $k_BT/e$, as opposed to 40-80 times of $k_BT/e$ needed for the current transistors. Here $k_B$ is the Boltzmann constant and T is the absolute temperature. That would make such devices less noisy and consuming less power.

However, the experimental results on YBS and similar devices reported to date remain illusive. In almost all of the efforts, such as those in T-branch[2], Y-junction[3,4,5], and ballistic rectifiers[6,7], the electrical characterization was carried out at a source-drain bias significantly larger than the thermal voltage. As a result, the transport mechanisms deviate too much away from the (near) equilibrium condition, which deems necessary to maintain the long coherence length. Therefore, the coherent transport plays no significant role in the characteristics of devices reported so far. Recently, Monte Carlo simulation on the above-mentioned devices[2-7] further indicates[8] that all of the main features reported, though supposedly from ballistic transport, can be described classically .

In this Letter, we report the first experimental demonstration of a Y branch switch in the quantum regime using InAs electron wave guides. All transport characteristics reported here are acquired with an excitation voltage less than the thermal energy. Thus the system is kept near equilibrium. Our YBS shows significant deviation from classical transport in their characteristics. These novel characteristics arise, to a great extent, because we have used InAs quantum wires in our YBS's. Here, we will discuss the fabrication, transistor characteristics, and qualitative explanation of the observed features.

Our transistors are built on InAs single quantum wells grown by molecular beam epitaxy. A typical structure has a 2 μm undoped GaSb buffer, GaSb/AlSb smoothing superlattice, a 100 nm AlSb bottom barrier and a 17 nm bare InAs quantum well (QW). The two-dimensional (2D) electron gas in InAs QWs can have a long mean free path[9] and long coherence length. In addition, the InAs/AlSb system has a number of properties that are advantageous for nanofabrication and for studying low-dimensional physics. First, the surface Fermi level pinning position in InAs is above its conduction band minimum, giving InAs conducting wires with nanometer widths.[10] Second, small InAs electron effective mass ($m_e^* = 0.023\ m_0$) gives large quantization energy.

Magnetotransport studies are first used to calibrate the 2D electron concentration ($n_{2D}$) and mobility ($\mu_{2D}$) of the as-grown sample. Quantum Hall plateaus and Shubnikov de Haas (SdH) oscillations, both characterizing 2D electrons, are clearly observed on photo-lithographically patterned Hall bars. We obtained an $n_{2D}$ of $3.09 \times 10^{12}$ cm$^{-2}$, $2.08 \times 10^{12}$ cm$^{-2}$ and $1.08 \times 10^{12}$ cm$^{-2}$, and a $\mu_{2D}$ of $1.06 \times 10^4$ cm$^2$/Vs, $1.67 \times 10^4$ cm$^2$/Vs and $1.54 \times 10^4$ cm$^2$/Vs, at 300 K, 77 K and 4K, respectively. The corresponding Fermi wavelength ($\lambda_F$), Fermi energy, and $l_e$, are calculated to be 14 nm, 140 meV, and 307 nm at 300K, 17nm, 114 meV, and 397 nm at 77K, and 24nm, 85 meV, and 264 nm at 4K, respectively. To independently verify the relatively long elastic mean free path in independent measurements, we have used electron-beam lithography and wet-etching to fabricate cross-junctions for measuring the "bend resistance."[11] At room temperature, 100nm cross junctions indeed display negative bend resistance[12] which can be destroyed by magnetic focusing in a perpendicular magnetic field, further confirming the ballistic transport.[13]

The operation of the YBS requires lateral electric field. We have adopted an in-plane gate structure,[3,14] in which two isolated, coplanar, conductive regions on the opposite sides of the Y junction are used as the two side-gates. When we apply bias to these two side gates differentially, a lateral electric field arises perpendicular to the symmetry axis of the Y-junction. In contrast, with the two side-gates shorted, a gate bias induces or depletes the electrons in the channel in the same way as a typical field-effect transistor does through capacitive coupling.

The practical difficulty in realizing YBS has been to fabricate electron wave guides in nanometer scale without depleting electrons in narrow channels. We have overcome this hurdle by using InAs material that has zero lateral depletion width. The fabrication procedures are discussed in the following. First, the large area bonding pads with leads are defined by photolithography, metal evaporation, and lift-off. Second, electron beam lithography is used to define the YBS patterns, as well as the two side-gates. Third, using the patterned electron beam resist as an etch-mask, wet chemical etching is subsequently applied to form a trench, 17nm deep and several tens of nanometer wide, which physically and electrically isolate the side gates from the conductive Y branches. Fig. 1 shows an atomic force micrograph of a YBS device. The Y junction is designed in such a way that the source (Source) and the two drains (Drain1 and Drain2) are all tapered toward the immediate junction area in order to avoid significant backscattering. The two side-gates (gate1 and gate2) are shaped to be with a sharp angle, aiming at localizing the lateral electric field in the junction region, thereby facilitating the observation of the mode switching effect.

In order to verify the gating effect and to obtain insight into our sample system, we have first characterized the QPC's by the standard lock-in technique at 4.2K. The two side-gates are shorted for a common-mode measurement. In the common-source configuration, the ac excitation voltage at the drain is kept at $30\mu V_{rms}$ to avoid the undesirable, distracting "self-gating" effect.[15] The drain current is measured with a transimpedance amplifier with a gain of 1Mohm or 10Mohm. The input impedance is less than 10ohms. The ac conductance is recorded as the gate voltage is swept from -1V to 1V. Figure 2 shows the typical current-voltage characteristics of a QPC at 4.2K and its numerical derivative. As expected, a series of plateaus, amplified as peaks in the derivative, is observed. We have identified the positions of the plateaus by including a series resistance $R_S$, i.e., the measured conductance can be expressed to be $G_N = 1/[1/(N \cdot \frac{2e^2}{h}) + R_S]$, where $N$ is the number of one-dimensional (spin unresolved) channels with different transverse quantization energies below the Fermi level, and $R_S$ is series resistance. As shown by dashed lines in Fig. 2, the calculated $G_N$ fits well the measured data with $R_S$=1150ohm. This successful observation of the quantized conductance not only verifies the gating effect, but also demonstrates that the fabricated devices, including the QPCs and the YBS's, are indeed shorter than the elastic mean free path at 4.2K.

Because of the ballistic, coherent transport, the characteristics of the gated YBS are expected to be drastically different from their classical counterparts, such as differential pair amplifiers. Figure 3(a) shows the transfer characteristics of our YBS with 76nm wide junction. There is no measurable gate leakage current (<< 1pA) in our measurements. The trans-conductance through Drain1 and, separately, Drain2 are shown as a function of the sweeping differential gate voltage: $-0.83V < V_{gate1} < 0.83V$, with $V_{gate1} = -V_{gate2}$. When $V_{gate1}$ ($= -V_{gate2}$) is swept, the electric field in the lateral direction steers the wave functions and the interference pattern of the injected electrons. Under such differential gating, the conductances through Drain1 and Drain2 show peaks and valleys, and these oscillatory features are anti-correlated. The sum and the difference of the two measured conductances are shown in Fig. 3(b) and 3(c), respectively. For $|V_{gate1}| < 0.25V$, there is little gating effect, and we attribute this lack of response to electrostatic screening by the InAs surface states. That is, the Fermi level pinning position must be shifted to allow for the gating effect to occur. The oscillatory trans-conductance is a manifestation of phase coherence in a multi-mode one-dimensional electron wave guide.

Using a model InAs quantum well structure with a thickness of 17nm, only the lowest 2D size-quantization level is occupied. Based on the 2D ground state, the confinement in the transverse lateral direction in a 100nm wide wire further defines a series of one-dimensional (1D) modes. A Fermi energy of 85meV suggests that there are seven 1D modes occupied. In the actual sample, there are about 4 to 9 1D modes populated under equilibrium at 4.2K.

Conductance through Drain1 in our device appears to be bigger than conductance through Drain 2, as shown in Fig.3(a). This is clearly a non-universal feature, which arises because of a given realization of confinement potential, impurities and surface roughness, and arrangement of contacts in the device. Such "mesoscopic" behavior is no surprise for devices with sizes smaller than the phase breaking length[16]. A given potential realization in the device defines the oscillatory features in conductances through Drain1 and Drain2. The scale of these features is a fraction of conductance through a single transverse channel. The sum of two conductances increases with increasing lateral electric field in either polarity.

We now qualitatively explain these features. The eigenenergies of the transverse 1D modes, $E_n$, the corresponding wave functions, $\eta_n(y)$, and the longitudinal component of the electron wave functions $\psi_{ny}(x)$ are defined by the given device potential. Here, $n$ labels the quantized states. Electrons in n-th transverse channel propagating in a wave guide from the source can be reflected back with amplitude $r_n$, transferred to Drain1 wave guide with amplitude $\varepsilon_{1n}$ and transferred to the Drain2 wave guide with amplitude $\varepsilon_{2n}$. In Y switch with symmetric branches and symmetric confinement with respect to the axis of the source wave guide, odd number transverse modes have maximum of the wave function on this axis. When scattering between transverse modes is ignored, the sum of conductances is given by $G_\Sigma = \sum_n \frac{2e^2}{h}\left(\varepsilon_{1n}^2 + \varepsilon_{2n}^2\right) = \sum_n \frac{2e^2}{h}(1 - r_n^2)$. If no transverse electric field is applied, the reflection amplitude in an odd channel is maximal because electrons are strongly scattered by a beam splitter that separates Drain1 from Drain 2. The corresponding contribution off odd-number channels to conductance is minimal. However, when a lateral electric field is applied, electrons are more likely to propagate into Drain1 or Drain 2, and are less scattered by the splitter. Therefore, reflection becomes weaker, and the sum of conductances through Drain1 and Drain 2 becomes bigger. For even modes, the scattering off the beam splitter would likely increase with electric field, the reflection amplitude will grow, and the corresponding contribution to conductance decreases. However, because in our structure the total number of modes is odd, the effect in the odd channels prevails, and the sum of condutances through the two drains increases by a fraction of the unit of conductance, as the data shows indeed. Another possible explanation of the observed behavior of $G_\Sigma$ is the increase of the number of conducting channels contributing to the current by one in the presence of transverse electric field. In parabolic-shaped confinement potential for the wave guide, e.g., the energies of all transverse states decrease with applied electric field, and it is possible that more transverse states will have total energy $E_F$, and contribute to conductance. However, e.g., for rectangular confinement, the quantization energy of the ground state decreases, but energies of all other levels increase, and no addition of channels is possible. This underscores the role of mesoscopic phase coherent effects and importance of specific realization of confinement (and impurity potential) for a given structure.

The observation that conductances through Drain1 and Drain2 are predominantly out of phase agrees well with the notion that electric field steers the wave functions from one drain into another. The individual Drain1 and Drain2 conductance characteristics is determined by the quasi-localized states formed around the beam splitter, i.e., the corresponding resonant transmission. Similar states affect electron transmission through T-filter structures.[17]

In conclusion, our observations provide strong evidence that the electron wave packet is being steered by an external electric field. The switching mechanism is quantum mechanical. That is, while the gate bias is coupled capacitively to the YBS region, the modulation of the drain current is not through depletion of the conductive path. We find our results unexpected from what is originally proposed.[1] Specifically, in the presence of multiple modes, we have observed oscillatory transconductance. Most strikingly, the oscillations measured from the two drains are always out of phase, in good agreement with our simple model. These results not only verify the quantum steering of electron wave functions in a semiconductor wave guide, but also open up possibilities for further studies of quantum switches in multiple-terminal, nanometer-scale structure for information processing.

Acknowlegement: This work is supported in part by LPS and ONR. CHY acknowledges discussion with Prof. P.T. Ho at UMCP, and Prof. Thylén and Dr. Forsberg at KTH.

Figure Captions

Fig. 1: Atomic force micrograph of a finished YBS, where the narrowest neck has a width of 76nm. The dark region is etched and the AlSb buffer is revealed. The light region is InAs quantum well. The terminals for electrical measurement are labeled.

Fig. 2: Two terminal source-drain conductance of a quantum point contact measured as a function of common-mode gate biasing. Within the shown gate bias range, there is no measurable dc gate leakage current (<< 1pA). The numerical derivative is shown in circles. The numbering of quantized conductance plateaus is shown.

Fig. 3: (A) Measured conductances from Drain1 and Drain2 are shown as a function of the differential gate voltage. (B) The sum of the two conductances. (C) Conductance of Drain 1 is subtracted by that of Drain2.

Figure 1 (Jones et al.)

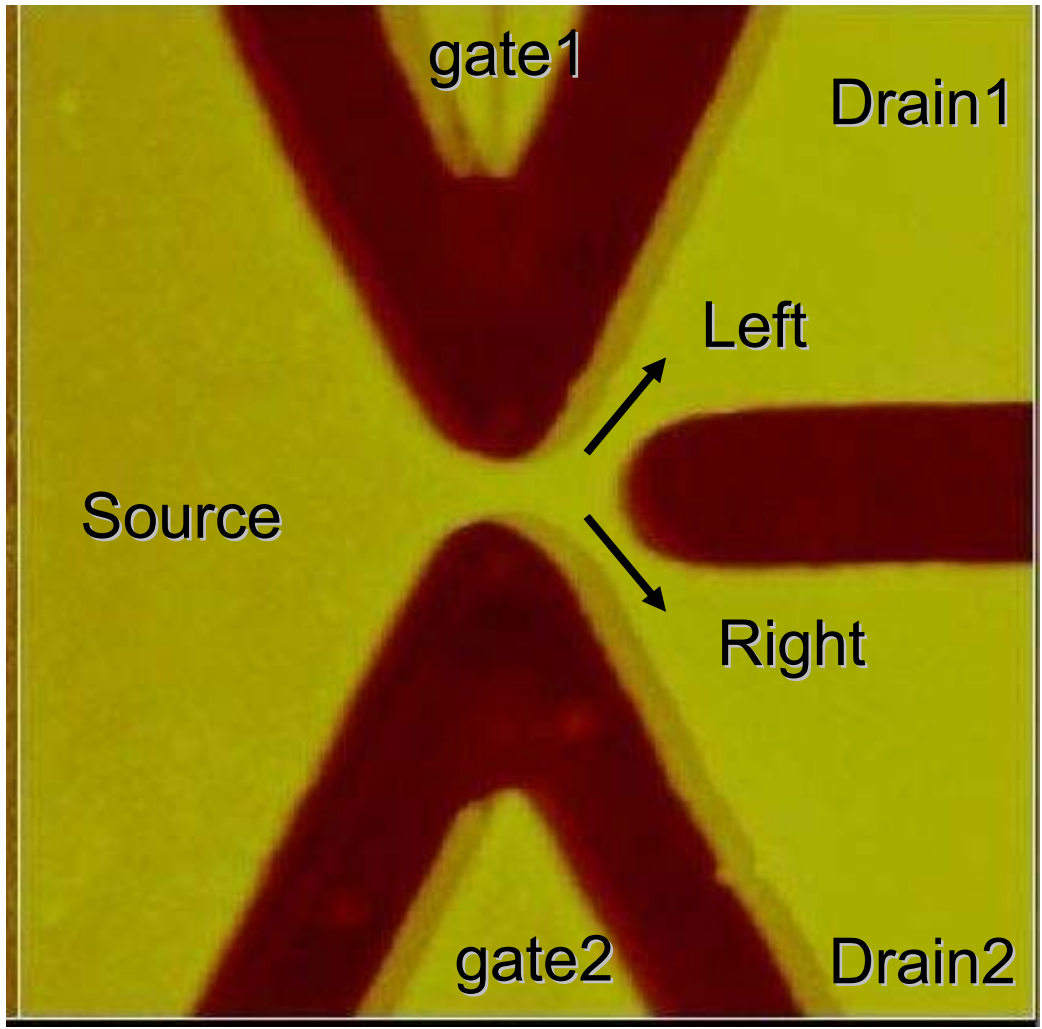

Figure2 (Jones, et al.)

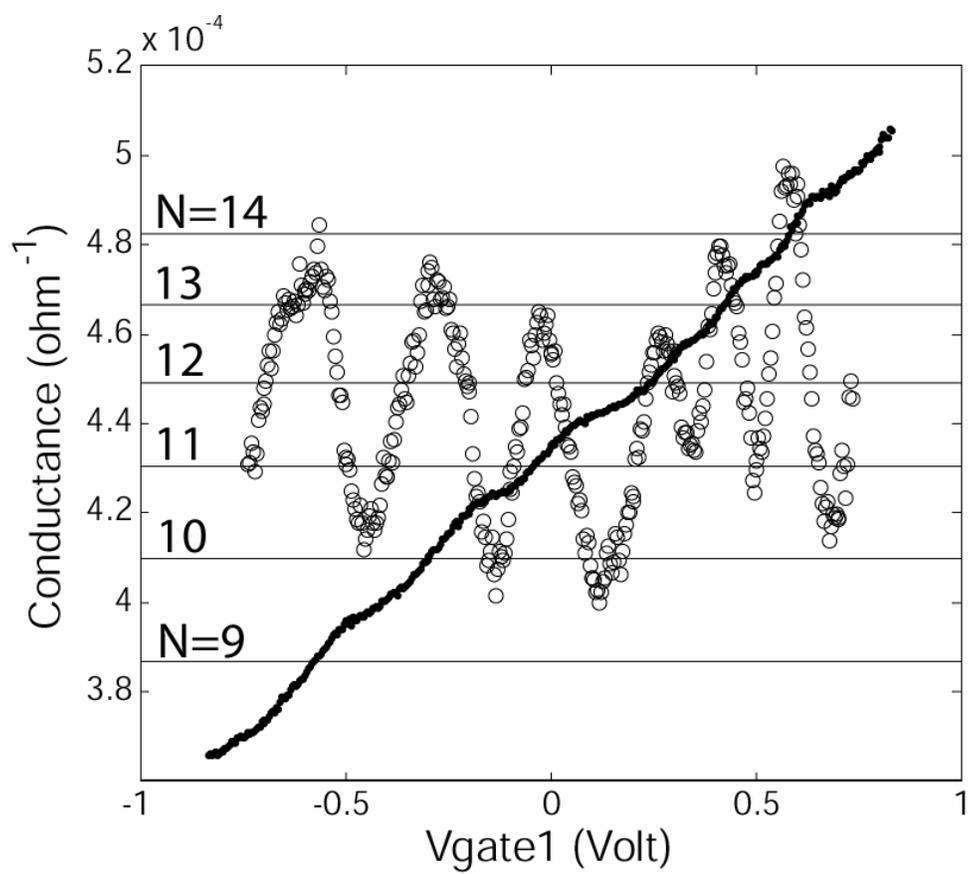

Figure3 (Jones et al.)

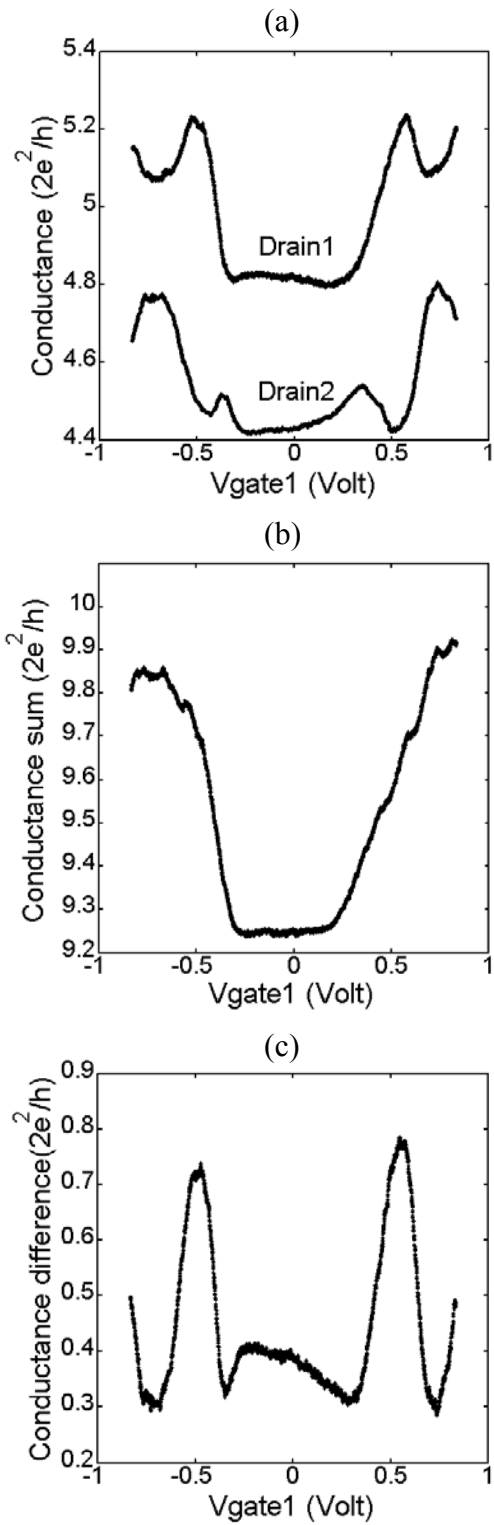

(a)

(b)

(c)

References

1 T. Palm and L. Thylén, App. Phys. Lett. 60, 237 (1992).

2 I. Shorubalko, H. Q. Xu, I. Maximov, P. Omling, L. Samuelson, and W. Seifert, "Nonlinear operation of GaInAs/InP-based three-terminal ballistic junctions," Appl. Phys. Lett., 79, 1384, 2001.

3 L. Worschech, B. Weidner, S. Reitzenstein, and A. Forchel, Appl Phys. Lett., 78, 3325, 2001.

4 L. Worschech, H. Q. Xu, A. Forchel, and L. Samuelson, "Bias-voltage induced asymmetry in nanoelectronic Y-branches," Appl Phys. Lett., 79, 3287, 2001.

5 K. Hieke and M. Ulfward, "Nonlinear operation of the Y-branch switch: ballistic switching mode at room temperature," Phys. Rev. B, vol. 62, pp. 16 727–16 730, 2000.

6 A. M. Song, A. Lorke, A. Kriele, J. P. Kothaus, W. Wegscheider, and M. Bichler, "Nonlinear electron transport in an asymmetric microjunction: a ballistic rectifier," Phys. Rev. Lett., 80, 3831, 1998.

7 A. M. Song, P. Omling, L. Samuelson, W. Seifert, I. Shorubalko, and H. Zirath, "Operation of InGaAs/InP-based ballistic rectifiers at room temperature and frequencies up to 50 GHz," Jpn. J. Appl. Phys., 40, L909–L911, 2001.

8 J. Mateos, B. G. Vasallo, D. Pardo, T. González, J.-S. Galloo, S. Bollaert, Y. Roelens, and A. Cappy, IEEE Tr. ED 50, 1897 (2003).

9 S. J. Koester, B. Brar, C. R. Bolognesi, E. J. Caine, A. Patlach, E. L. Hu, and H. Kroemer, Phys. Rev. B **53**, 13063 (1996).

10 K. A. Cheng, C. H. Yang, and M. J. Yang, Appl. Phys. Lett. 77, 2861 (2000); J. Appl. Phys. 88, 5272 (2000); T. H. Chang, C. H Yang, and M. J. Yang, Physica Status Solidi B 224, 693 (2001).

11 G. Timp, H.U. Baranger, P. deVegvar, J.E. Cunningham, R. E. Howard, R. Behringer, and P.M. Mankiewich, Phys. Rev. Lett. 60, 2081 (1988).

12 G.M. Jones, M.J. Yang, and C.H. Yang, to be submitted.

13 C. W. J. Beenakker and H. van Houten, Quantum Transport in Semiconductor Nanostructures, in Solid State Physics, v 44, Eds. H. Ehrenreich and D. Turnbull, (Academic Press, Boston 1991).

14 K.K. de Vries, P. Stelmaszyk, and A.D. Wieck, J. Appl. Phys. 79 8087 (1996).

15 S. Reitzenstein, L. Worschech, P. Hartmann, M. Kamp, and A. Forchel, Phys. Rev. Lett. 89, 226804 (2002).

16 Mesoscopic Phenomena in Solids, Vol. eds. B.L. Altshuler, P.A. Lee and R.A. Webb. (North-Holland, Amsterdam, New York, 1991).

17 A.A. Kiselev and U. Roessler, Semiconductor Science and Technology, **11** 203 (1996).